\documentclass[12pt]{iopart}
\usepackage{graphicx}% Include figure files
\usepackage{dcolumn}% Align table columns on decimal point
\usepackage{bm}% bold math

\def\la{\mathrel{\mathpalette\fun <}}
\def\ga{\mathrel{\mathpalette\fun >}}
\def\fun#1#2{\lower3.6pt\vbox{\baselineskip0pt\lineskip.9pt
  \ialign{$\mathsurround=0pt#1\hfil##\hfil$\crcr#2\crcr\sim\crcr}}}

\begin{document}

%%%%%%%%%%%%%%%%%%%%%%%%%%%%%%%%%%%%%%%%%%%%%%%%%%%%%%%%%%%%%%%%%%%%%%
%% Front Material %%%%%%%%%%%%%%%%%%%%%%%%%%%%%%%%%%%%%%%%%%%%%%%%%%%%
%%%%%%%%%%%%%%%%%%%%%%%%%%%%%%%%%%%%%%%%%%%%%%%%%%%%%%%%%%%%%%%%%%%%%%

\title[Cosmological Gravitational Wave Background from Phase Transitions in 
Neutron Stars]{Cosmological Gravitational Wave Background from Phase Transitions in Neutron Stars}

\author{G\"unter Sigl\dag}

\address{\dag
APC~\footnote[2]{UMR 7164 (CNRS, Universit\'e Paris 7, CEA, Observatoire de Paris)} (AstroParticules et Cosmologie),
11, place Marcelin Berthelot, F-75005 Paris, France\\
GReCO, Institut d'Astrophysique de Paris, C.N.R.S.,
98 bis boulevard Arago, F-75014 Paris, France}

\begin{abstract}
  It has recently been suggested that collapse of neutron stars induced
  by a phase transition to quark matter can be a considerable source of  
  gravitational waves with kHz frequencies. We demonstrate that if about one    
  percent of all neutron stars undergo this process, the resulting cosmological
  gravitational wave background would reach about $10^{-10}$ times the
  critical density. The background would peak at kHz frequencies and could
  have an observationally significant tail down to Hz frequencies.
  It would be comparable
  or higher than other astrophysical backgrounds, for example, from
  ordinary core collapse supernovae, from r-mode instabilities in
  rapidly rotating neutron stars, or from magnetars. The scenario is
  consistent with cosmological backgrounds in neutrinos and photons.
\end{abstract}



\maketitle

%%%%%%%%%%%%%%%%%%%%%%%%%%%%%%%%%%%%%%%%%%%%%%%%%%%%%%%%%%%%%%%%%%%%%%
%% Introduction %%%%%%%%%%%%%%%%%%%%%%%%%%%%%%%%%%%%%%%%%%%%%%%%%%%%%%
%%%%%%%%%%%%%%%%%%%%%%%%%%%%%%%%%%%%%%%%%%%%%%%%%%%%%%%%%%%%%%%%%%%%%%

\section{Introduction}

A phase transition from nuclear to quark matter within a neutron
star (NS) with millisecond scale rotation period can induce a collapse
that releases up to $\sim10^{53}\,$erg
in energy, which potentially relates this process to violent phenomena
such as core collapse supernovae~\cite{Yasutake:2004kx} or
$\gamma-$ray bursts (GRBs)~\cite{Olinto:1986je,Cheng:1995am,Bombaci:2000cv}.
Recent numerical simulations suggest that up to a few percent of this energy
may be released as gravitational waves (GWs)~\cite{Lin:2005zd}.

If a significant fraction of NSs undergo such a phase transition
induced collapse, the energy released will contribute to cosmological
backgrounds in photons, neutrinos, and GWs. In the present paper we
estimate these backgrounds and point out that the stochastic GW
background may be comparable to GW backgrounds from various other
astrophysical sources. We specifically compare the power and statistical
properties of this potential background with the ones from
standard core collapse supernovae~\cite{Buonanno:2004tp},
NS-NS coalescence~\cite{Schneider:2000sg,Regimbau:2005tv},
r-mode instabilities in NSs with millisecond 
periods~\cite{Owen:1998xg,Ferrari:1998jf}, and from 
magnetars~\cite{Regimbau:2005ey},
as well as from various processes in the early universe.
We also find a significant constraint on the fraction of the total
energy released in such phase transitions in form of MeV $\gamma-$rays.

We use natural units with $\hbar=c=1$ throughout.

\section{The Source Mechanism}

We first recall that the energy radiated in GWs per frequency interval for an
individual event at distance $D$, $(dE_{\rm gw}/df)(f)$, is related to
the Fourier transform $\tilde{h}(f)\equiv\int_{-\infty}^{+\infty}dt\,e^{-i2\pi ft}h(t)$ of the dimensionless strain amplitude $h(t)$ by
\begin{equation}
\frac{dE_{\rm gw}}{df}(f)=\frac{16\pi^2 D^2}{15G_{\rm N}}\,|f\tilde{h}(f)|^2\,,
\label{Egw}
\end{equation}
where $G_{\rm N}$ is Newton's constant.

We model the strain in the phase transition scenario as
\begin{equation}
  h(t)\propto\exp\left(-\Gamma t\right)\sin f_0t\quad\hbox{for}\quad t\geq0
  \,,\label{h}
\end{equation}
which roughly reflects the form found in the simulations in
Ref.~\cite{Lin:2005zd}. This leads to a Fourier transform
\begin{equation}
  |\tilde{h}(f)|^2\propto\frac{f_0^2}{(f_0^2-f^2+\Gamma^2)^2+4\Gamma^2f^2}
  \,,\label{tilde_h}
\end{equation}
Note that the total energy emitted in GWs $E_{\rm gw}$ from Eq.~(\ref{Egw})
converges for this spectral form. We use $f_0=2\,$kHz, $\Gamma=1/3\,$ms for
the damping scale which is typical for the most rapidly rotating
progenitors in Ref.~\cite{Lin:2005zd}, and normalize such that
$E_{\rm gw}\simeq2\times10^{51}\,$erg, the more optimistic case from Ref.~\cite{Lin:2005zd}.

Most of the energy released in the phase transition is actually carried
away in form of photons and neutrinos, whose total energy we will denote
by $E_{\rm tot}\ga(1-5)\times10^{52}\,$erg. This is roughly the energy required to power a $\gamma-$ray burst. The gravitational wave energy is
typically $E_{\rm gw}\simeq0.05E_{\rm tot}$~\cite{Lin:2005zd}.

%%%%%%%%%%%%%%%%%%%%%%%%%%%%%%%%%%%%%%%%%%%%%%%%%%%%%%%%%%%%%%%%%%%%%%
\section{Cosmological Event Rates}
%%%%%%%%%%%%%%%%%%%%%%%%%%%%%%%%%%%%%%%%%%%%%%%%%%%%%%%%%%%%%%%%%%%%%%

\begin{figure}[t]
\includegraphics[width=0.8\textwidth,clip=true]{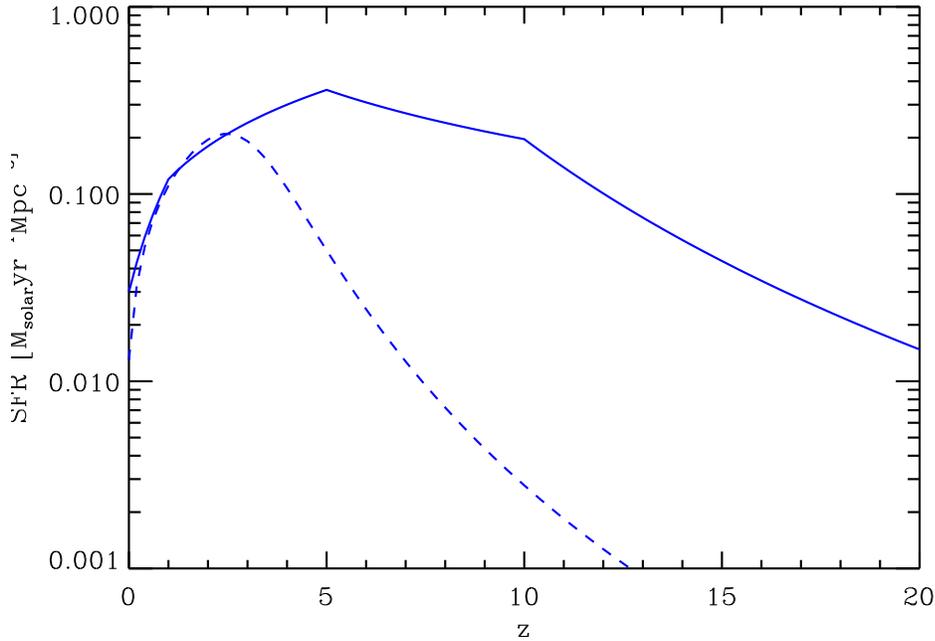}
\caption{Models for the star formation rate as function of redshift.
The solid line is from Ref.~\cite{Bromm:2005ep} which includes a putative PopIII
component, whereas the dashed line respresents the Baldry \& Glazebrook
form~\cite{Baldry:2005df} fitted in Ref.~\cite{Hopkins:2006bw}.}
\label{fig1}
\end{figure}

The cosmic star-formation rate (SFR) and, as a consequence, the formation
rate $R(z)$ of NSs is reasonably well known at redshifts
$z\la5$~\cite{Daigne:2004ga,Hopkins:2006bw}. Especially for $z\la1$
it is strongly constrained by the Super-Kamiokande limit on the
electron antineutrino flux from cosmological core collapse
supernovae~\cite{Hopkins:2006bw,Malek:2002ns}. In contrast, the SFR is poorly
known for $z\ga5$. We use the fits shown in Fig.~\ref{fig1}.

We are interested in the number of events per unit mass. This number
is given by an expression of the general form
\begin{equation}
  \lambda\equiv\chi\frac{\int_{M_{\rm min}}^{M_{\rm max}}dm\xi(m)}
  {\int dm\,m\xi(m)}\,,\label{fraction}
\end{equation}
where we neglect any redshift dependence.
Here, $\xi(m)$ is the mass function, usually taken to be of the
Salpeter form $\xi(m)\propto m^{-2.35}$ between $0.1M_\odot$ and
$100M_\odot$, $[M_{\rm min},M_{\rm max}]$ is the progenitor
mass range over which NSs form, and the integral in the denominator
goes over all stellar masses. Further, $\chi$ is the fraction
of NSs eventually undergoing the phase transition under
consideration. The event rate $R(z)$ is then given by the
product of the SFR and $\lambda$.

For NSs, $M_{\rm min}\simeq10M_\odot$, $M_{\rm max}\simeq40M_\odot$,
so that $\lambda\simeq5\times10^{-3}\,\chi\,M_\odot^{-1}$.
Note that $\lambda\propto M_{\rm min}^{-1.35}$ and could be significantly 
larger if NSs are also formed from progenitors with masses
$m<10M_\odot$. The above value for $\lambda$ also roughly corresponds
to today's rate of core collapse 
supernovae which is $\simeq2\times10^{-4}\,{\rm Mpc}^{-3}\,{\rm yr}^{-1}$
which is to be expected given that most of the NSs are born in type II
supernova events. Most of the hot young NSs are born with 10-20 ms rotation
periods due to processes such as magnetic braking. We assume that 
$\chi\simeq1\%$ of the NSs undergo a phase transition. This is roughly the 
estimated fraction of NSs which are born with millisecond scale
rotation periods~\cite{Woosley:2006ie,Heger:2004qp}, required to trigger
the phase transition. The same sub-population of NSs
would be relevant for GW emission between $\sim100\,$Hz and $\sim2\,$kHz
from r-mode instabilities in rapidly rotating NSs, in contrast to Ref.~\cite{Owen:1998xg,Ferrari:1998jf} which assumed $\chi\sim1$.

Another possible channel for phase transitions to occur in NSs
is in low-mass X-ray binaries where the NS is spun up by accretion
from the low mass companion~\cite{Cheng:1995am,Cheng:1997af}. The
formation efficiency of such systems is hard to estimate, but
typical numbers are $\chi\sim10^{-3}-10^{-2}$~\cite{Pfahl:2003uc},
similar to the estimated fraction of millisecond NSs born in core
collapse supernovae.

Even larger values for $\lambda$ could be motivated if these phase transitions
are connected to short hard $\gamma-$ray bursts. A best estimate
of $\sim 10^{-4}\,{\rm Mpc}^{-3}\,{\rm yr}^{-1}$ for the comoving rate of
faint short hard $\gamma-$ray bursts at zero redshift has recently been 
given~\cite{Nakar:2005bs}. Comparing with  Fig.~\ref{fig1}, this corresponds to 
$\lambda\sim5\times10^{-3}\,M_\odot^{-1}$, which would correspond to
values for $\chi$ comparable to unity.

For the NS-NS coalescence and magnetar scenarios, $\lambda$ is actually
given by the same expression Eq.~(\ref{fraction}), with estimated fractions
$\chi=0.3\%$~\cite{Regimbau:2005tv} and $\chi=8\%$~\cite{Regimbau:2005ey}
of all NSs contributing, respectively.

We will neglect any time delay between star formation and NS formation
because the lifetime $\sim10^8\,$yr of $\ga10\,M_\odot$ NS progenitors
is short compared to the Hubble time.

\section{The Gravitational Wave Background}

For simplicity, we assume that all NSs have identical emission
characteristics.

The energy density in GWs at frequency $f$ per logarithmic frequency
interval in units of the cosmic critical density $\rho_{\rm c} =
3H_0^2/(8 \pi G_{\rm N})$ can be written as~\cite{Phinney:2001di}
\begin{equation}
\label{los}
\Omega_{\rm gw}(f)=\frac{1}{\rho_{\rm c}}\int_0^\infty dz\,
\frac{R(z)}{1+z}\left|\frac{dt}{dz}\right|
f_z\frac{dE_{\rm gw}}{df}(f_z)\,,
\end{equation}
where $R(z)$ is the event rate per comoving volume,
$f_z\equiv(1+z)f$. The cosmological model enters with
$|dt/dz|=[(1+z)H(z)]^{-1}$ and, for a flat geometry,
\begin{equation}
\label{cosmo}
H(z)= H_0\left[\Omega_{\rm M}(1+z)^3+\Omega_{\Lambda}\right]^{1/2}\,.
\end{equation}
We will use the parameters $\Omega_{\rm M}=0.3$,
$\Omega_{\Lambda}=0.7$, and $H_0=h_0\,100~{\rm km}~{\rm s}^{-1}~{\rm
Mpc}^{-1}$ with $h_0=0.72$.

For a stochastic GW background the dimensionless strain
power $\left|f\tilde{h}(f)\right|^2$ is related to the dimensionless
GW energy density in Eq.~(\ref{los}) by~\cite{Phinney:2001di}
\begin{equation}
  h_0^2\,\Omega_{\rm gw}(f)=\frac{2\pi^2h_0^2}{3H_0^2}f^2
  \left|f\tilde{h}(f)\right|^2=6.25\times10^{35}
  \left(\frac{f}{{\rm Hz}}\right)^2\,\left|f\tilde{h}(f)\right|^2
  \,.\label{Omega_h}
\end{equation}

\begin{figure}[t]
\includegraphics[width=0.8\textwidth,clip=true]{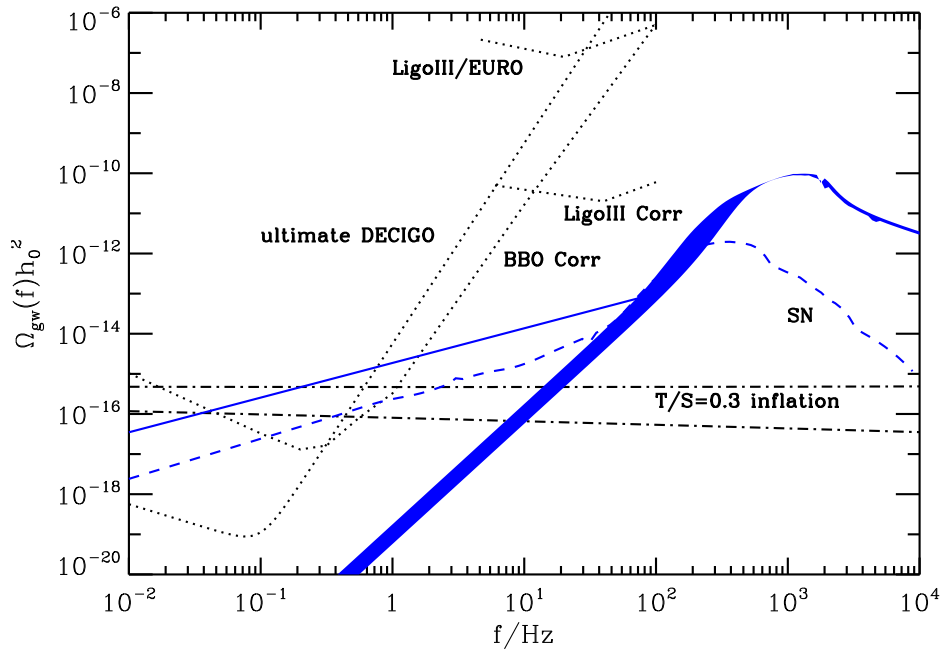}
\caption{Spectrum of GW background from Eq.~(\ref{los}) assuming
about $\chi=1\%$ of all NSs undergo a phase transition, releasing
$E_{\rm gw}\sim2\times10^{51}\,$erg in GWs. The band represents the uncertainty
related to the different SFRs shown in Fig.~\ref{fig1}. The solid line
is the maximal low-frequency tail due to anisotropic neutrino emission.
For comparison, the dashed line is the maximal background from
conventional type II supernovae discussed in Ref.~\cite{Buonanno:2004tp}.
Further, the horizontal dash-dotted lines represent a maximum version
of the GW stochastic spectrum produced during slow-roll inflation
assuming  $T/S=0.3$ for the ratio of the
tensorial and scalar contributions to the cosmic microwave background
radiation anisotropy and $\pm 10^{-3}$ for the running of the tensorial power-law index~\cite{Turner:1996ck}. The dotted lines show approximate 
sensitivities of  the ground based interferometer LigoIII/EURO and LigoIII
in correlation mode~\cite{Buonanno:2003th}, and of possible second generation
space-based interferometers such as the Big Bang Observatory 
(BBO)~\cite{Seto:2005qy} and the ultimate DECIGO~\cite{Seto:2001qf},
as indicated.}
\label{fig2}
\end{figure}

Fig.~\ref{fig2} shows the resulting GW background for the NS phase transition
scenario, the band representing uncertainties due to the different
SFRs from Fig.~\ref{fig1}, but not due to other quantities such as
$E_{\rm gw}$ and $\chi$ which were fixed to fiducial values,
as indicated. Note that GW 
densities are proportional to $\lambda$ from
Eq.~(\ref{fraction}), and thus also to the fraction $\chi$
of NSs subject to the phase transition. They are
also proportional to the average total GW energy output $E_{\rm gw}$.
As a result, the largest uncertainties come from the parameters
$\chi$ and $E_{\rm gw}$, whereas uncertainties due to the SFR are
moderate because the background is dominated by redshifts $z\la3$
where the SFR is reasonably well known, see Fig.~\ref{fig1}.

The stochastic GW background from cosmological supernovae was studied
recently in Ref.~\cite{Buonanno:2004tp}, and its optimistic
estimate is reproduced in Fig.~\ref{fig2}. While the event rate $R(z)$
used there is higher than in the present scenario (since supposedly
only a percent fraction of core collapse supernovae likely give rise to
NSs rotating rapidly enough to undergo phase transitions),
the individual source signal is very different:
In the phase transition scenario, the signal from one event
at kHz frequencies is much
higher than the one due to convection in the simulations in 
Ref.~\cite{Mueller:2003fs} and, as a consequence, the total energy
emitted in GWs is much larger, 
$\sim10^{-3}\,M_\odot$ versus $\sim10^{-8}\,M_\odot$.
As a result, the background from phase transitions can be comparable or
higher (above $\sim$100 Hz) to the background from conventional
type II supernovae.

There could also be an enhanced low frequency tail if
the strain $h(t)$ converges to a non-vanishing constant for
$t\to\infty$ due to anisotropic neutrino
emission~\cite{Epstein:1978dv, Turner:1978jj}.
For $f\ll\,$kHz, $dE_{\rm gw}/df\simeq15G_{\rm N}\,
(E_\nu q)^2$~\cite{Buonanno:2004tp},
where $E_\nu\leq E_{\rm tot}$ is the total energy emitted in neutrinos,
and $|q|\la1$ is the average dimensionless quadrupole. Whereas $E_\nu$ is
about an order of magnitude smaller than $E_\nu\sim3\times10^{53}\,$erg
in standard type II supernovae, the anisotropy could be much larger
than the order percent anisotropy expected in hot NSs without
phase transition~\cite{Mueller:2003fs}. This is
because the NS would oscillate strongly and be highly deformed after the
phase transition. In one of the simulations in Ref.~\cite{Lin:2005zd},
for example, the ratio of the polar to equatorial radius is $\simeq0.7$.
As in core collapse supernovae, the neutrinos would likely be partially
trapped and emitted from a neutrinosphere because the release of
$\sim5\times10^{52}\,$erg in internal energy would heat up the NS to
$\sim10-20\,$MeV.

We can parametrize the low-frequency tail as
\begin{equation}
  \Omega_{\rm gw}(f)\sim3.5\times10^{-15}\,
  \left(\frac{E_\nu q}{5\times10^{52}\,{\rm erg}}\right)^2
  \left(\frac{\lambda}{5\times10^{-5}\,M_\odot^{-1}}\right)
  \left(\frac{f}{{\rm Hz}}\right)\,.\label{low-f}
\end{equation}
The maximum of this tail, corresponding to the fiducial values in
Eq.~(\ref{low-f}) is shown in Fig.~\ref{fig2} and reflects
the uncertainty in the low-frequency signal: In the absence of a
"GW memory effect", $\tilde{h}(f)\to\,$const, see Eq.~(\ref{tilde_h}) and thus
$\Omega_{\rm gw}(f)\propto f^3$ for $f\to0$, see Eqs.~(\ref{Egw})
and~(\ref{los}). In contrast,
in the presence of burst with memory $f\tilde{h}(f)\to\,$const and
$\Omega_{\rm gw}(f)\propto f$ for $f\to0$.

\begin{figure}[t]
\includegraphics[width=0.8\textwidth,clip=true]{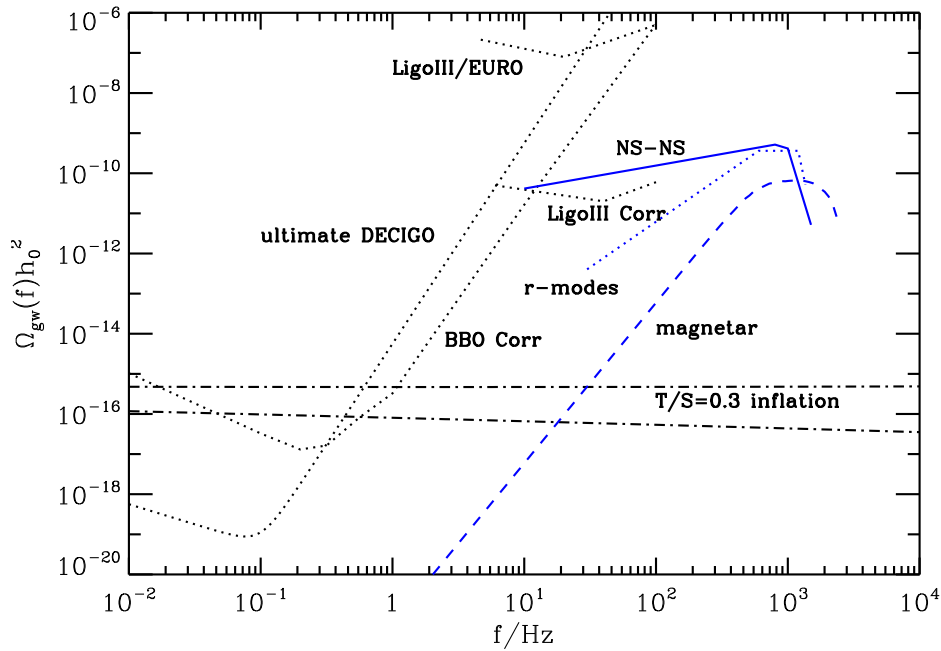}
\caption{GW backgrounds from various other astrophysical mechanisms for 
comparison to Fig.~\ref{fig1}. The solid line represents that part of the 
NS-NS coalescence signal from Ref.~\cite{Regimbau:2005tv} which has a 
"popcorn"-character, similar to the phase transition scenario considered here, 
see below. The dotted line corresponds to GWs from r-mode instabilities emitted
by the fastest rotating NSs~\cite{Owen:1998xg,Ferrari:1998jf},
assumed to constitute a fraction $\sim1\%$ of all NSs.
The dashed line shows the magnetar scenario~\cite{Regimbau:2005ey}.}
\label{fig3}
\end{figure}

Fig.~\ref{fig3} shows various other astrophysical backgrounds for
comparison: The solid line shows a possible popcorn like signal from NS-NS
coalescence~\cite{Regimbau:2005tv}. We note that other work used a lower
maximal frequency which would cut off the signal above 
$\sim200\,$Hz~\cite{Schneider:2000sg}. The dotted line shows a stochastic 
background from r-mode instabilities emitting GWs between
$\sim100\,$Hz and $\sim1.4\,$kHz. It has been corrected by the small fraction
$\chi\sim1\%$ of all NSs nowadays believed to be born with sufficiently small,
millisecond scale rotation periods, in contrast to the original
Refs.~\cite{Owen:1998xg,Ferrari:1998jf} which assumed that a substantial
fraction of all NSs are born with rotation periods close to maximal.
Finally, the dashed line represents the magnetar 
scenario~\cite{Regimbau:2005ey}. The NS phase transition scenario could thus 
produce a background which is comparable to or larger than
these backgrounds both at kHz frequencies and below $\sim10\,$Hz.

Finally, the background from NS phase transitions shown in Fig.~\ref{fig2}
would potentially mask several possible
GW relic backgrounds from the early universe, at least at frequencies
between $\simeq0.1\,$Hz and $\sim1\,$Hz, where it becomes close to
gaussian and thus indistinguishable to other gaussian backgrounds, see
next section.
Such early universe backgrounds include the one from standard inflation
whose maximum is shown in Fig.~\ref{fig2}, but also more speculative
ones such as from quintessential inflation which predicts
$\Omega_{\rm gw}h_0^2\sim6\times10^{-16}(f/{\rm Hz})$ above
$\sim10^{-3}\,$Hz~\cite{Peebles:1998qn}.

%%%%%%%%%%%%%%%%%%%%%%%%%%%%%%%%%%%%%%%%%%%%%%%%%%%%%%%%%%%%%%%%%%%%%%
\section{Statistical Properties of Gravitational Wave Background and 
Detectability}
%%%%%%%%%%%%%%%%%%%%%%%%%%%%%%%%%%%%%%%%%%%%%%%%%%%%%%%%%%%%%%%%%%%%%%

The event rate as seen from Earth is
\begin{equation}\label{rate}
\int_{0}^{\infty}dz\,\frac{R(z)}{1+z}\frac{dV}{dz}=
\int_{0}^{\infty}dz\,R(z)\frac{4\pi r^2(z)}{(1+z)H(z)}\,,
\end{equation}
where $dV/dz$ is the fractional volume element, the cosmic expansion
rate at redshift $z$ is given by Eq.~(\ref{cosmo}), and $r(z)$ is the
comoving coordinate, $dr=(1+z)dt$.

The duty cycle can be estimated by multiplying the integrand
of Eq.~(\ref{rate}) with $(1+z)t_{\rm coh}[(1+z)f]$, where
$t_{\rm coh}(f)$ is the timescale over which the
frequency $f$ is emitted coherently. A probably more realistic
estimate is obtained by weighting the resulting integral over redshift with
the redshift dependent contribution to the GW signal given by the
integrand in Eq.~(\ref{los}).

\begin{figure}[t]
\includegraphics[width=0.8\textwidth,clip=true]{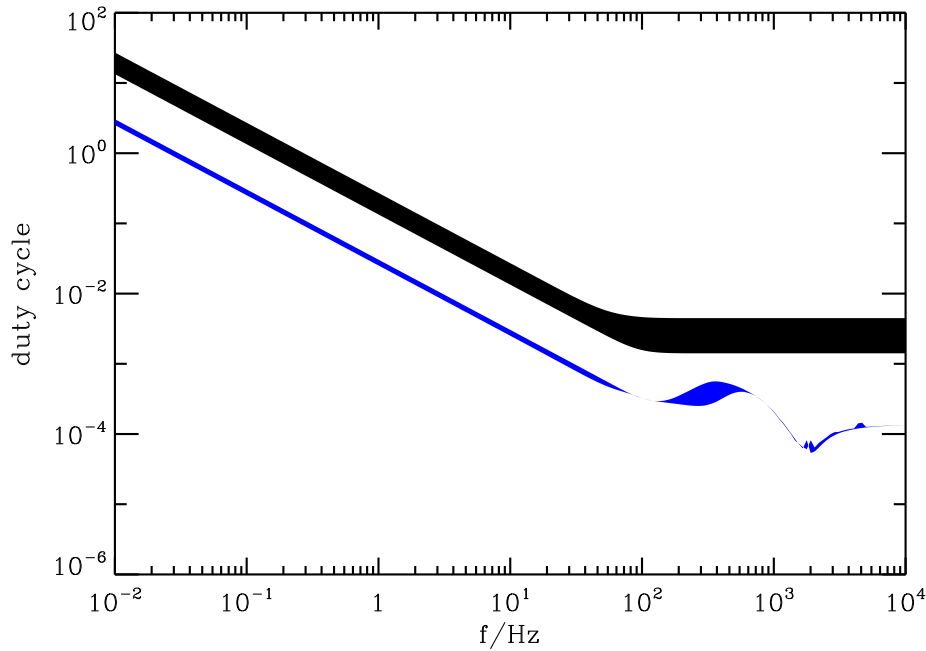}
\caption{The duty cycle as function of frequency for the NS phase transition
scenario. The blue (lower) band includes weighting with the GW signal,
whereas the dark (upper) band does not. Again, the bands represent
the uncertainty related to the different SFRs shown in Fig.~\ref{fig1}.}
\label{fig4}
\end{figure}

Fig.~\ref{fig4} shows the duty cycle obtained by these two estimates
for the NS phase transition scenario for which
$t_{\rm coh}(f)\simeq\max(1/\Gamma,1/f)$. If it is smaller than
one, the duty cycle roughly corresponds to the variable $\xi$ in
Ref.~\cite{Drasco:2002yd}. The background is thus of
"popcorn" character. Note that the duty cycle is
proportional to $\lambda$. The minimal detectable GW energy density
$\Omega_{\rm gw}$ for a popcorn signal is comparable or, for
duty factors smaller than a few percent, smaller by factors of
a few compared to the minimal detectable background for gaussian 
noise~\cite{Drasco:2002yd}. Thus for
$\lambda\sim5\times10^{-3}\,M_\odot^{-1}$, the signal would
have a duty factor of a few percent and may thus
be marginally detectable by third
generation interferometers in correlation mode, see Fig.~\ref{fig2}.
This would require, however, that a fraction of order one of NSs
would have to contribute, which is unlikely. Also note that
microwave cavity detectors could reach sensitivities of order
$\Omega_{\rm gw}\sim10^{-9}$ at a few hundred
Hz~\cite{Bernard:2001kp,Maggiore:1999vm}.

The most advanced interferometers planned for the future such as
GEO-HF~\cite{danzmann} can reach sensitivities around $f^{1/2}\tilde{h}(f)
\sim5\times10^{-24}\,{\rm Hz}^{-1/2}$ in the $1-10\,$kHz regime.
One can easily deduce from Eqs.~(\ref{Egw}) and~(\ref{tilde_h}) that for
a source at distance $D$ (small compared to the Hubble radius), the maximum
of $f^{1/2}\tilde{h}(f)$ occurs at $f\simeq2\,$kHz with
$f^{1/2}\tilde{h}(f)\simeq3\times10^{-23}\,(D/10\,{\rm Mpc})^{-1}\,
{\rm Hz}^{-1/2}$ for the phase transition scenario. Such events would
thus be detectable out to
$\sim10\,$Mpc, with a rate $\simeq0.2\,(\chi/0.01)(D/10{\rm Mpc})^3
\,{\rm yr}^{-1}$.

At $f\sim\,$kHz the signal from core collapse supernovae
has a duty cycle slightly higher than the one of the phase transition
signal if $\chi\sim1\%$ of the NSs contribute to the phase transition
signal. This is because the supernova rate is then roughly a factor 100
higher, whereas the coherence time $t_{\rm coh}(f)\simeq1/f$ for
the incoherent process of convection is slightly
shorter than the one of the oscillations
right after the phase transition, for which $t_{\rm coh}(f)\simeq3\,$ms.

For both the r-mode instability~\cite{Owen:1998xg,Ferrari:1998jf} and 
magnetar~\cite{Regimbau:2005ey} scenarios, $t_{\rm coh}(f)\gg1/f$,
and the signal is Gaussian at all observable frequencies.
The NS-NS coalescence signal consists of a gaussian, a popcorn, and
a shot noise component~\cite{Regimbau:2005tv}.
A fraction of $\chi\sim0.003$ of all NSs are in suitable binaries to
contribute to the coalescence signal. As a consequence, the background
has statistical properties similar to the one for NS phase transitions.
The total power from NS-NS coalescence shown in Fig.~\ref{fig3} can
be comparable or somewhat higher than for the NS phase transition
scenario because the gravitational energy released is higher, 
$E_{\rm gw}\sim5\times10^{52}\,$erg~\cite{Regimbau:2005tv}.

%%%%%%%%%%%%%%%%%%%%%%%%%%%%%%%%%%%%%%%%%%%%%%%%%%%%%%%%%%%%%%%%%%%%%%
\section{Electromagnetic and Neutrino Emissions}
%%%%%%%%%%%%%%%%%%%%%%%%%%%%%%%%%%%%%%%%%%%%%%%%%%%%%%%%%%%%%%%%%%%%%%

The energy going into GWs during the phase transitions considered in
Ref.~\cite{Lin:2005zd} represents only a fraction of a
few percent of the total energy released $E_{\rm tot}$.
Fig.~\ref{fig2} shows that an energy
$E_{\rm gw}\simeq2\times10^{51}\,$erg released per event in GWs and
a number of events per unit mass $\lambda\simeq5\times10^{-5}M_\odot^{-1}$
corresponds to $\Omega_{\rm gw}\sim10^{-10}$,
quite independent of the poorly known high redshift
evolution of the SFR. By simple scaling, this implies an energy density
\begin{equation}
\Omega_{\nu+\gamma} h_0^2\simeq3\times10^{-9}\,
  \left(\frac{E_{\rm tot}}{5\times10^{52}\,{\rm erg}}\right)\,
  \left(\frac{\lambda}{5\times10^{-5}M_\odot^{-1}}\right)\label{O_tot}
\end{equation}
is released in form of photons and/or neutrinos.
Photons and neutrinos are expected to be emitted on timescales
$\ga10\,{\rm s}\gg1/\Gamma$ than GWs, a factor $\ga3\times10^3$
larger than GWs which are emitted on a timescale $\simeq1/\Gamma\simeq3\,$ms.
Fig.~\ref{fig4} then implies that these backgrounds are quasi-continuous for
$\lambda\ga5\times10^{-5}\,M_\odot^{-1}$, provided the emissions are
not strongly beamed.

This emission can certainly not be predominantly in form of MeV $\gamma-$rays
because the cosmological diffuse background of photons above an MeV
has an energy density $\Omega_{\rm MeV}h_0^2\simeq4\times10^{-10}$.
Put another way, the fraction $f_{\rm MeV}$ that can be released in these events
is constrained by
\begin{equation}
  f_{\rm MeV}\la0.1\,
  \left(\frac{E_{\rm tot}}{5\times10^{52}\,{\rm erg}}\right)^{-1}
  \left(\frac{\lambda}{5\times10^{-5}M_\odot^{-1}}\right)^{-1}\,.\label{f_MeV}
\end{equation}

Since the energy density in lower energy backgrounds are larger, no significant
constraint on the fraction going into photons of energy below $\sim100\,$keV
result.

Obviously, if only a very small fraction of the SFR contributes to
these events, i.e. if $\lambda$ is small, and/or the total energy
released $E_{\rm tot}$ is small, this constraint
becomes weak or disappears. However, the fiducial values for $\lambda$ and $E_{\rm tot}$ used here are not unrealistic, and imply significant constraints 
on emission into MeV $\gamma-$rays.

If phase transitions are to be associated with GRBs, a fraction
\begin{equation}
  f_{\rm MeV}\sim10^{-3}\,
  \left(\frac{E_{\rm tot}}{5\times10^{52}\,{\rm erg}}\right)^{-1}
  \left(\frac{\lambda}{5\times10^{-5}M_\odot^{-1}}\right)^{-1}
\end{equation}
released as $\sim\,$MeV $\gamma-$rays would in fact be sufficient
because GRB emissions correspond to a (non-continuous) photon
background with energy density
$\Omega_{\rm GRB}\sim2\times10^{-12}$. Since $\sim10^3$ GRBs are
visible per year, this would imply a beaming factor
\begin{equation}
  \sim3\times10^3\left(\frac{\lambda}{5\times10^{-5}M_\odot^{-1}}\right)\,.
\end{equation}
The corresponding Lorentz factor would be roughly the square root
of this.

%%%%%%%%%%%%%%%%%%%%%%%%%%%%%%%%%%%%%%%%%%%%%%%%%%%%%%%%%%%%%%%%%%%%%%
\section{Conclusions}                          \label{sec:conclusions}
%%%%%%%%%%%%%%%%%%%%%%%%%%%%%%%%%%%%%%%%%%%%%%%%%%%%%%%%%%%%%%%%%%%%%%
We have estimated the cosmological background of gravitational waves
for the scenario where a fraction $\chi\sim1\%$ of all neutron stars
are born as millisecond
pulsars and undergo a phase transition to a quark star, releasing
$\sim10^{51}\,$erg in gravitational waves, as suggested by recent
simulations. In terms of the corresponding event rate per unit stellar mass,
$\lambda$, this background would have a duty factor of 
$\sim5\times10^{-4}\left[\lambda/(5\times10^{-5}\,M_\odot^{-1})\right]$ above
$\sim50\,$Hz and
$\sim0.03\left[\lambda/(5\times10^{-5}\,M_\odot^{-1})\right](f/{\rm Hz})^{-1}$ 
below $\sim50\,$Hz.
However, most of the energy flux is at kHz frequencies, a factor 10-100
above the maximum sensitivity of ground based interferometers.
Using statistics specialized to detecting popcorn type noise with
collocated aligned detectors at the technological limit, the background may be 
marginally detectable around 100 Hz for
$\lambda\sim5\times10^{-3}\,M_\odot^{-1}$ which is only possible if
the majority of all neutron stars would be born with millisecond periods
and undergo a phase transition. In this unrealistic case the background
would have duty factors of a few percent.

The total energy
in gravitational waves can be, however, quite substantial even for
moderate fractions $\chi$ of neutron stars contributing. In units
of the critical density one obtains
$\Omega_{\rm gw}h_0^2\sim10^{-10}\left[E_{\rm gw}/(2\times10^{51}
\,{\rm erg})\right]\left(\chi/0.01\right)$. In addition, since neutrinos
tend to be emitted very anisotropically in the form of jets, there
could be a low-frequency tail of order
$\Omega_{\rm gw}(f)\la10^{-15}\,
\left[\lambda/(5\times10^{-5}\,M_\odot^{-1})\right](f/{\rm Hz})$,
likely larger than for ordinary type II supernovae. A low-frequency tail
of that size would mask the maximal gravitational background in ordinary
inflation models between $\sim0.1\,$Hz and $\sim1\,$Hz where it
would also be gaussian.
At these frequencies this tail could be detectable by future space based
interferometer projects such as BBO or DECIGO.

The gravitational wave background from phase transitions in rapidly
rotating newly born neutron stars can thus be comparable or higher
than other astrophysical backgrounds.
Detection of such a signal could provide insight into the nature of
compact objects.

If most of the energy released in the phase transition is in form of
neutrinos, the resulting diffuse MeV neutrino flux would constitute
$\sim10^{-3}(\chi/0.01)$ of the flux due to standard type II core
collapse supernovae, and would thus be consistent with existing upper
limits~\cite{Malek:2002ns}.
The fraction of the energy released in form of MeV $\gamma-$rays
is constrained by observed diffuse backgrounds to be less than
$\sim10\%(\chi/0.01)^{-1}$. A fraction $\sim10^{-3}(\chi/0.01)^{-1}$
released into the band of $100\,{\rm keV}-1\,$MeV $\gamma-$rays
would suffice to account for $\gamma-$ray bursts.

%%%%%%%%%%%%%%%%%%%%%%%%%%%%%%%%%%%%%%%%%%%%%%%%%%%%%%%%%%%%%%%%%%%%%%
%% Acknowledgments %%%%%%%%%%%%%%%%%%%%%%%%%%%%%%%%%%%%%%%%%%%%%%%%%%%
%%%%%%%%%%%%%%%%%%%%%%%%%%%%%%%%%%%%%%%%%%%%%%%%%%%%%%%%%%%%%%%%%%%%%%

\ack

We thank Fr\'ed\'eric Daigne and Elisabeth Vangioni-Flam for informative discussions. We especially
thank Alessandra Buonanno, Thomas Janka, and Georg Raffelt for discussions
and very useful comments on an early version of the manuscript.

\section*{References}

%%%%%%%%%%%%%%%%%%%%%%%%%%%%%%%%%%%%%%%%%%%%%%%%%%%%%%%%%%%%%%%%%%%%%%
%% Bibliography %%%%%%%%%%%%%%%%%%%%%%%%%%%%%%%%%%%%%%%%%%%%%%%%%%%%%%
%%%%%%%%%%%%%%%%%%%%%%%%%%%%%%%%%%%%%%%%%%%%%%%%%%%%%%%%%%%%%%%%%%%%%%

\end{document}